%% file: Radpour_WCNC2024.tex
\documentclass[conference]{IEEEtran}
\IEEEoverridecommandlockouts
\usepackage{amsmath,amssymb,amsfonts}
\usepackage{graphicx}
\usepackage{textcomp}
\usepackage{xcolor}
\usepackage[nolist,nohyperlinks]{acronym}
\usepackage{lipsum}
\usepackage{booktabs} 

\usepackage{algorithm2e}

\usepackage{float}
\usepackage{nicefrac}
\usepackage{thmtools}
\usepackage{enumerate}
\usepackage[caption=false,font=footnotesize]{subfig}

\def\BibTeX{{\rm B\kern-.05em{\sc i\kern-.025em b}\kern-.08em
    T\kern-.7667em\lower.7ex\hbox{E}\kern-.125emX}}

\def\mathlette#1#2{{\mathchoice{\mbox{#1$\displaystyle #2$}}%
                           {\mbox{#1$\textstyle #2$}}%
                           {\mbox{#1$\scriptstyle #2$}}%
                           {\mbox{#1$\scriptscriptstyle #2$}}}} 
\renewcommand{\Vec}[1]{\mathlette{\boldmath}{#1}} 

\DeclareMathOperator*{\argmax}{arg\,max}

\begin{document}

\title{Active Reconfigurable Intelligent Surface for the Millimeter-Wave Frequency Band: Design and Measurement Results}
\author{Hamed Radpour$^\ast$, Markus Hofer$^\ast$, Lukas Walter Mayer$^\dagger$, Andreas Hofmann$^\dagger$, Martin Schiefer$^\dagger$ \\ and Thomas Zemen$^\ast$\\
	$^\ast$AIT Austrian Institute of Technology, Vienna, Austria\\
    $^\dagger$Siemens Aktiengesellschaft Oesterreich, Vienna, Austria\\
	Email: hamed.radpour@ait.ac.at}
	
	
\maketitle
\input{abbrev}

\begin{abstract}
Reconfigurable intelligent surfaces (RISs) will play a key role to establish reliable low-latency millimeter wave (mmWave) communication links for indoor automation and control applications. In case of a blocked line-of-sight  between the base station (BS) and the user equipment (UE), a RIS mounted on a wall or on a ceiling enables a bypass for the radio communication link. In this work, we present an active RIS for the mmWave frequency band. 
Each RIS element uses a field effect transistor (FET) to amplify the reflected signal and an orthogonal polarization transformation to increase the isolation between impinging and reflected radio wave. By switching the bias voltage at gate and drain of the FET we can establish four states for each RIS element: two reflection states with different phase shifts, an active amplification and an off state. We present results of the active RIS with 37 patch antenna elements arranged in a hexagonal grid for a center frequency of 25.8 GHz. The RIS field patterns obtained by numerical simulations and by empirical measurements in an anechoic chamber are compared. They show a good match and the received power is improved by 12 dB in the active mode of the RIS compared to the reflective mode.
\end{abstract}


\section{Introduction}
The reconfigurable intelligent surfaces (RIS) technology enables an adaptation of the radio communication environment to specific needs of an application. The RIS is composed of an array of small sub-wavelength patch elements with adjustable reflection coefficients. The reflection coefficients can be set such, that the wave impinging on the RIS is reflected to a desired direction or focused at a defined point in space \cite{Basar19}. 

As RISs enable intelligent control of wireless communication, they can be used for improving coverage and reliability of the communication link \cite{khawaja2020coverage, Rains23}. In this paper we are specifically interested to use the RIS in an indoor automation and control scenario with high-reliability and low-latency requirements for the mmWave communication link. The RIS shall provide a bypass in case the line-of-sight is blocked between base station (BS) and user equipment (UE).

Passive RIS prototypes have been designed to enhance the signal-to-noise-ratio (SNR) at the UE side for center frequencies below 6\,GHz \cite{Rains23} and in the mmWave band at 28\,GHz \cite{Dai20}. In \cite{Bjornson20a} the authors analyze the required surface area for a passive RIS that is needed to establish a reflected BS-RIS-UE link that is as strong as the one achieved by a decode and forward relay. In \cite[(22)]{najafi2020physics} the required number of passive reflective antenna elements is analyzed that is needed to establish a reflected BS-RIS-RX link that is as strong as a hypothetical direct link from BS to UE. For both objectives, a large number of RIS elements is required for outdoor scenarios, e.g., in \cite{tang21} a passive RIS working at $10.5\,$GHz with $10200$ elements ($1\,\text{m}\times 1\,$m size) has been shown that is suitable for BS-RIS and RIS-UE distances of about $100\,$m. Such a large number of RIS elements requires larger computational complexity for the RIS control algorithm, increases the RIS cost and its physical size.

The number of required RIS elements can be reduced by introducing active RIS elements. An active RIS element includes an amplifier that introduces an adjustable amplification and phase shift. The authors of \cite{zhi2022active} show that an active RIS outperforms a passive RIS under the same power budget. Hence, active RIS elements can either increase the SNR at the UE when the number of RIS elements is kept constant, or the number of RIS elements can be reduced for a constant SNR target helping to reduce the cost and size of the RIS. A smaller RIS is also advantageous for time-sensitive automation and control applications that are the focus of this paper.  

In \cite{Zhang23} a fractional programming algorithm for solving the optimization problem of joint beam-forming and reflect precoding for an active RIS is presented. A single element reflect-amplifier is demonstrated for a center frequency of $2.3$\,GHz, but no integration into a complete active RIS is shown in \cite{Zhang23}. A $2\times2$ active RIS using a compact micro-strip reflect amplifier for each RIS element for a center frequency of $5\,$GHz is shown in \cite{rao23}.

In \cite{wu2022wideband} an active RIS is presented that uses the orthogonal polarization transform, i.e., the incoming signal is received, e.g., with horizontal polarization, amplified and transmitted with vertical polarization. This setup increases the isolation of the impinging and the reflected wave to suppress self oscillation of the active RIS element. The orthogonal polarization transforms is demonstrated in \cite{wu2022wideband} with a $2\times2$ antenna array connected to a single amplifier that can operate in a wide frequency range from $5\,$GHz to $6\,$GHz.

Summarizing, so far active RISs with only four RIS elements have been demonstrated operating at sub-$6\,$GHz frequencies. Therefore, an active RIS for the mmWave frequency band and a control algorithm are necessary for future reliable and low-latency 6G wireless communication systems.

\subsection*{Scientific Contributions}
\begin{itemize}
	\item We present an active RIS for an indoor industrial environment for the the mmWave frequency band containing $37$ elements. Each RIS element contains a field effect transistor (FET) and can be set into four states: two reflective states, one amplification state and an off state. In the active state the RIS elements use a polarization transform to increase the isolation between impinging and reflected wave. Up to the best of our knowledge, this is the first active RIS working at mmWave frequencies.
	\item We perform measurements of the active RIS in an anechoic chamber, to verify and evaluate the performance in terms of received signal power as well as the beam pattern resolved in azimuth and elevation. These measurements are compared to numerical simulations using the path loss model from \cite{tang22} that is extended to the active RIS case. 
    \item The comparison between the reflective and active mode shows a signal power gain of $12\,$dB at the position of the UE.
\end{itemize}

\section{Active RIS Path Loss Model}
\begin{figure}[ht!]
	\centering
	\includegraphics[width=\columnwidth]{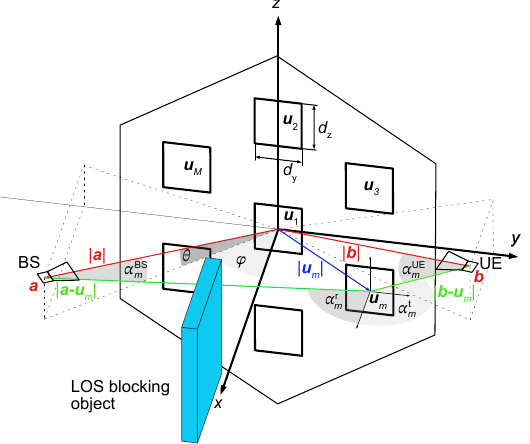}
	\caption{RIS coordinate system for a hexagonal RIS element placement in the yz-plane. The BS horn antenna radiates from position $\Vec{a}$ towards the center of the RIS at $\Vec{0}=(0,0,0)$ over a distance of $|\Vec{a}|$, similarly the UE horn antenna at position $\Vec{b}$ is within a distance of $|\Vec{b}|$ pointing towards the origin. The LOS is blocked between BS and UE. The picture is not to scale to improve clarity.} 
	\label{fig:RIScoordinates}
\end{figure}

A path loss model for a passive RIS with rectangular elements is presented in \cite{tang21} for below 6\,GHz and a refined model for the mmWave frequency range in \cite{tang22}. We are considering that there is no direct link between BS and UE and there is a line-of-sight (LOS) channel for the BS-RIS and the RIS-UE link. The BS is located at $\Vec{a}=(a_\text{x},a_\text{y},a_\text{z})$ in Cartesian coordinates. We also use spherical coordinates $\tilde{\Vec{a}}=(a_r, a_\varphi, a_\theta)$ depending on the context. We measure azimuth $a_\varphi$ between the x-axis and the projection of the vector $\tilde{\Vec{a}}$ on the x-y plane and elevation $a_\theta$ between the projection in the x-y plane and the vector itself. The radius $a_\rho=|\tilde{\Vec{a}}|$, see Fig. \ref{fig:RIScoordinates}. 

The UE is located at $\Vec{b}$ and the $M$ RIS element center points at $\Vec{u}_m$ with $m\in\{1,\ldots, M\}$. The received power at the UE position is given as
\begin{multline} 
P_{\text{UE}}=\underbrace{P_{\text{BS}}\frac{G_{\text{BS}} G_{\text{UE}}(d_\text{y} d_\text{z})^2}{16\pi ^{2}}}_C\\
\times\left | \sum_{m=1}^{M}\Gamma_{m}\underbrace{\frac{\sqrt{F_{m}^{\text{c}}}e^{-j2\pi (|\Vec{a}-\Vec{u}_m|+|\Vec{b}-\Vec{u}_m|)/\lambda}}{|\Vec{a}-\Vec{u}_m||\Vec{b}-\Vec{u}_m|}}_{D_m} \right |^{^{2}}.
\label{eq:ReceivedPower}
\end{multline}
where $P_{\text{BS}}$, $G_{\text{BS}}$, $G_{\text{UE}}$, $d_\text{y}$, and $d_\text{z}$ are the transmit power of the BS, the BS and UE antenna gain as well as the effective RIS element dimension in $y$ and $z$ direction, respectively \cite{tang22}. The complex reflection coefficient of each RIS element $m$ is denoted by $\Gamma_{m}$. The wavelength $\lambda=c_0/f$, where $f$ denotes the center frequency, and $c_0$ the speed of light. The combined antenna pattern of the BS antenna, the RIS element $m$ for receive and transmit operation as well as the UE antenna is described by
\begin{multline}
F^\text{c}_m=F^\text{BS}(\alpha^\text{BS}_m) F(\alpha^\text{r}_m) F(\alpha^\text{t}_m) F^\text{UE}(\alpha^\text{UE}_m)=\\
=\cos(\alpha^\text{BS}_m)^{\frac{G_\text{BS}}{2}-1}\cos(\alpha^\text{r}_m) \cos(\alpha^\text{t}_m) \cos(\alpha^\text{UE}_m)^{\frac{G_\text{UE}}{2}-1} =\\
=\left(\frac{|\Vec{a}|^2+|\Vec{a}-\Vec{u}_m|^2-|\Vec{u}_m|^2}{2|\Vec{a}||\Vec{a}-\Vec{u}_m|}\right)^{\frac{G_\text{BS}}{2}-1}\left(\frac{a_\text{x}}{|\Vec{a}-\Vec{u}_m|}\right)\\
\times \left(\frac{b_\text{x}}{|\Vec{b}-\Vec{u}_m|}\right)
\left(\frac{|\Vec{b}|^2+|\Vec{b}-\Vec{u}_m|^2-|\Vec{u}_m|^2}{2|\Vec{b}||\Vec{b}-\Vec{u}_m|}\right)^{\frac{G_\text{UE}}{2}-1}.
\end{multline} 
following \cite[(7)]{tang22} and the geometric configuration in Fig. \ref{fig:RIScoordinates}. Here $F^\text{BS}(\cdot)$, $F^\text{UE}(\cdot)$, and $F(\cdot)$ denote the rotation symmetric antenna pattern with respect to the main propagation direction for BS, UE, and RIS element, respectively. The angle with respect to the main wave propagation direction of each antenna is measured by $\alpha^\text{BS}_m$, $\alpha^\text{UE}_m$, $\alpha^\text{r}_m$ and $\alpha^\text{t}_m$ for the BS horn antenna, the UE horn antenna, the RIS element in reception and the RIS element in transmit direction, respectively.

\section{Reflection Coefficients}
\label{sec:RISControlAlgorithm}
We represent a complex number $\gamma\in\mathbb{C}$ in polar coordinates, and write $\gamma= \rho e^{j\xi}=(\rho, \angle \frac{\xi}{2\pi}360^\circ)$, where $\rho=|\gamma|$ is the absolute value of $\gamma$ and $\xi$ is the phase. A closed form solution to compute the phase $\xi_m$ of the reflections coefficients $\Gamma_m=\rho_m e^{j\xi_m}$ to maximize $P_\text{UE}$ is given in \cite[(12)]{tang21} as
\begin{equation}
\xi_m=\mod\left(\frac{2\pi (|\Vec{a}-\Vec{u}_m|+|\Vec{b}-\Vec{u}_m|)}{\lambda}, 2\pi\right) \in \mathbb{R}.
\label{eq:ReflectionPhase}
\end{equation}
assuming coordinates of BS, UE and RIS elements positions $\Vec{a}$, $\Vec{b}$ and $\Vec{u}_m$ are known. 

\begin{figure}
	\centering
	\includegraphics[width=0.9\columnwidth]{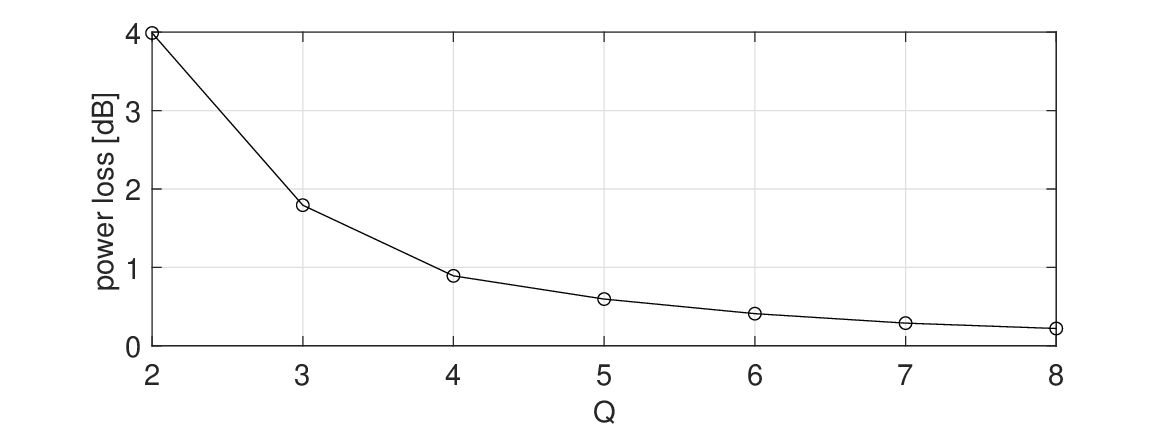}
	\caption{Loss in received power vs. the number of quantization intervals $Q$ for RIS parameters in Tab. \ref{tab:ActiveRIS}.}
	\label{fig:PowerLoss} 
\end{figure}

Quantizing $\xi_m$ into $Q$ equal intervals
\begin{equation}
\xi'_{m}=\frac{2\pi}{Q}\left(\left\lfloor\frac{\xi_m Q}{2\pi}\right\rfloor+0.5\right).
\label{eq:UniformQuantisation}
\end{equation}
simplifies the RIS element hardware implementation and reduces its cost. In Fig. \ref{fig:PowerLoss}  we plot the loss in received power at the UE location vs. the number of quantization intervals $Q$, evaluating \eqref{eq:ReceivedPower} numerically for the RIS parameters in Tab. \ref{tab:ActiveRIS}.
We can see that $Q=2$ quantization intervals cause a loss of $4\,$dB. Hence, a single bit quantization enables a simple RIS element realization with an acceptable loss in received power.

\begin{figure}
	\centering
	\includegraphics[width=\columnwidth]{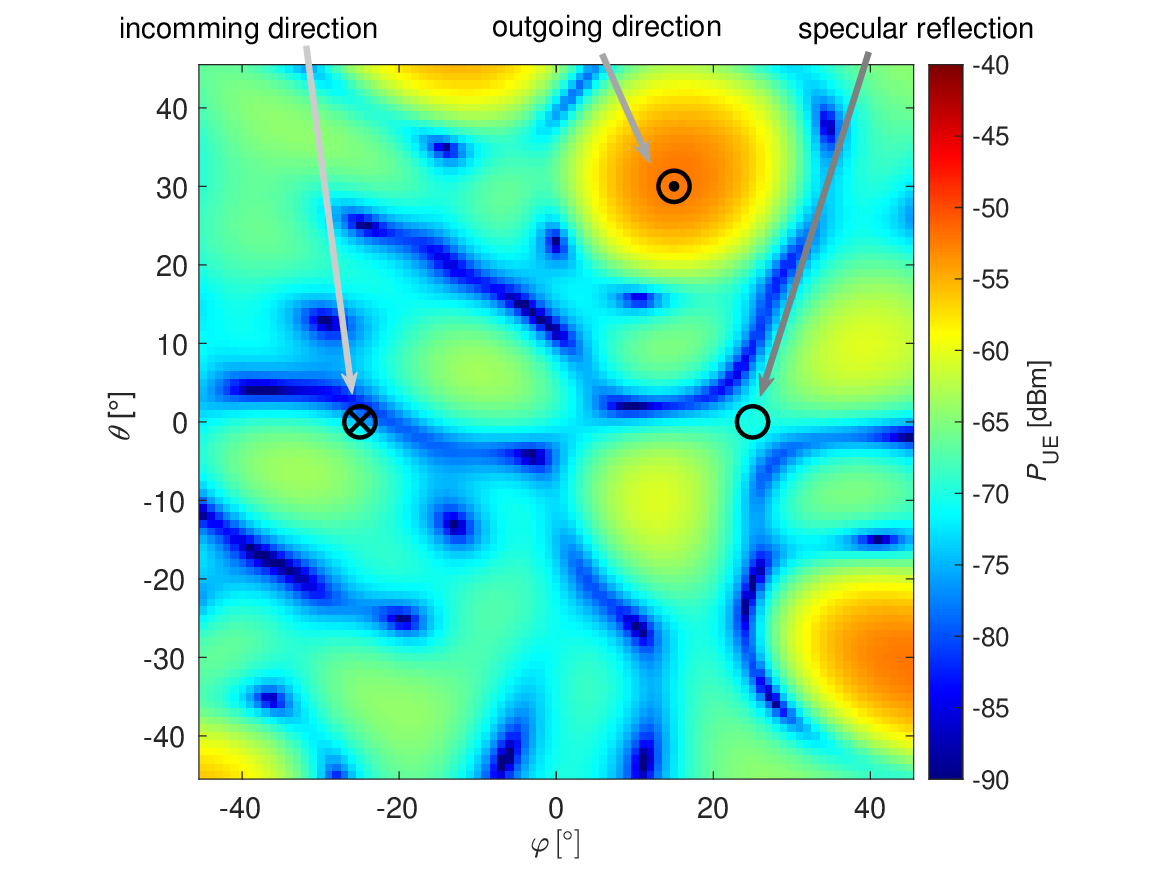}
	\caption{Numerical simulation result for the received power $P_\text{UE}(\varphi, \theta)$ vs. azimuth $\varphi$ and elevation $\theta$ for an ideal RIS with reflection coefficient alphabet $\mathcal{A}=\{(0.4, \angle 90^\circ) , (0.4, \angle 270^\circ) \}$. The BS is located at $\tilde{\Vec{a}}= ( 1.7\,\text{m}, -25^\circ, 0^\circ)$ and the signal shall be focused at the UE position $\tilde{\Vec{b}}=(1.7\,\text{m}, 15^\circ, 30^\circ)$.}
	\label{fig:ReflectiveIdeal}
		\vspace{-4mm} 
\end{figure}

For the mmWave application in indoor scenarios the obtained power at the UE position is the crucial parameter that we want to optimize in non-LOS situations. For $Q=2$ we obtain a reflection coefficient alphabet $$\Gamma_{m}\in\mathcal{A}=\{(0.4, \angle 90^\circ) , (0.4, \angle 270^\circ) \}.$$ We plot the resulting beam pattern in Fig. \ref{fig:ReflectiveIdeal}. The gain of $0.4$ is chosen to enable a direct comparison with the results in Section \ref{sec:Comparison}. 

In Fig. \ref{fig:ReflectiveIdeal} the radio signal is directed in the intended direction at the UE position $\tilde{\Vec{b}}=(1.7\,\text{m}, 15^\circ, 30^\circ)$ and the specular reflection is nicely suppressed. In the lower right corner a side lobe is visible that is caused by the $0.75\lambda$ RIS element spacing and the finite size of the RIS. 

In Section \ref{sec:RISHW} the active RIS element realization in our prototype hardware is explained. It will become clear that a practical reflection coefficient alphabet will deviate from the choice used for the simulation in Fig. \ref{fig:ReflectiveIdeal}, i.e. the elements of $\mathcal{A}$ will have a phase difference that deviates from $360^\circ/Q$. Hence, for a general $\mathcal{A}$ the closed form solution using \eqref{eq:ReflectionPhase} and \eqref{eq:UniformQuantisation} is not applicable. Therefore, we need to resort to the maximization of the received power according to 
\begin{equation}
    \argmax_{\Gamma_m \in \mathcal{A}\, \forall\, m\in\mathcal{I}_M}\left | \sum_{m=1}^{M}\Gamma_{m} D_m \right |^2.
\label{eq:MaximumReceivedPower}
\end{equation}
The solution of \eqref{eq:MaximumReceivedPower} requires the search over $|\mathcal{A}|^M$ possible solutions which is infeasible already for moderate values of $M$ as in Tab. \ref{tab:ActiveRIS}. Hence, we approximate \eqref{eq:MaximumReceivedPower} with a Monte Carlo search as defined in Algorithm \ref{alg:MCsearch}.
\RestyleAlgo{ruled}
\begin{algorithm}
\caption{Maximize received power for finite set of reflection coefficients.}
\label{alg:MCsearch}
\KwIn{ $\Vec{a}$, $\Vec{b}$, $\Vec{u}_m$, $\mathcal{A}$}
$\mathcal{I}_M=\{1,\ldots,M\}$\;
$\Gamma_{m}=0 \quad\forall\quad m\in\mathcal{I}_M$\;
\For{$i=1$ \KwTo $500$ }{
choose $m'\in\mathcal{I}_M$ randomly\;
$\argmax_{\Gamma_{m'}\in \mathcal{A}}\left | \sum_{m\in\mathcal{I}_M}\Gamma_{m} D_m \right |^2$\;
    }
\KwOut{$\Gamma_m, m\in\mathcal{I}_M$}
\end{algorithm}

We will use Algorithm \ref{alg:MCsearch} in Section \ref{sec:RISbeampattern} for the numerical simulations and empirical measurements of our hardware prototype. The algorithm selects RIS elements randomly and chooses the reflection coefficient from the finite alphabet $\mathcal{A}$ to maximize the received power at the UE. For the RIS configuration in this paper 500 iterations are sufficient. 

\section{Active RIS Element Design}
\label{sec:RISHW}
\begin{figure}
	\centering
	\includegraphics[width=.9\columnwidth]	{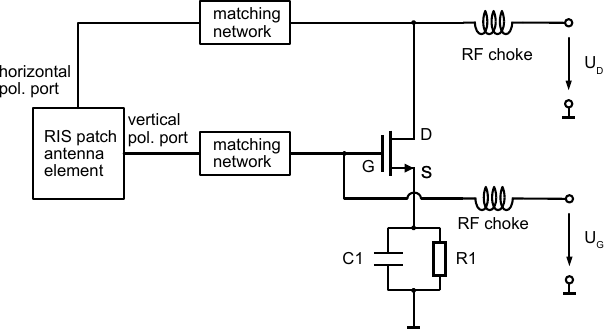}
	\caption{Active RIS element circuit diagram.} 
	\label{fig:RISCiruitDiagram}
\end{figure}

The RIS consists of $M$ micro-strip patch antenna elements on a multilayer printed circuit board (PCB). The RIS elements are arranged on a hexagonal grid with a spacing of $0.75 \lambda$. Each RIS patch antenna element has two ports, one for horizontal polarization and another one for vertical polarization. We exploit the attenuation between these two orthogonal antenna polarizations to obtain sufficient isolation between input and output of the amplifier at each RIS element. The amplification is realized with a single FET per RIS element. In Fig. \ref{fig:RISCiruitDiagram} we depict the simplified circuit diagram of a single active RIS element. 

The state of each RIS element $m$ is configured by two bits that switch between two possible bias voltages for the gate $U_{\text{G},m}\in\{u_1,u_2\}$ and the drain $U_{\text{D},m}\in\{u_3,u_4\}$ resulting in different reflection coefficient $\Gamma_m$. Hence, we use $U_{\text{D},m}$ to switch between reflective and active mode, and $U_{\text{G},m}$ to change the phase shift in reflective mode. For the reflective mode $\mathcal{A}_\text{R}=\{\gamma_1, \gamma_2\}$ where $|\gamma_1|,|\gamma_2|<1$ and for the active mode $\mathcal{A}_\text{A}=\{\gamma_3, 0\}$ where $|\gamma_3|>1$.  

The RIS elements have the following reflection coefficients in reflective mode:
$$\mathcal{A}_\text{R}=\{(0.4, \angle 0^\circ) , (0.4, \angle 67^\circ) \}.$$ In active mode: 
$$\mathcal{A}_\text{A}=\{(2, \angle 0^\circ) , ( 0, \angle 0^\circ) \}.$$
The RIS design is optimized for the active mode, hence the intended phase shift of $180^\circ$ for $\gamma_2$ could not be achieved. In active mode the RIS elements are either switched on or off, hence only about one half of all elements will contribute to the reflected signal, see Fig. \ref{fig:RISconfig}. The effects of these practical limitations will be analyzed below in Section \ref{sec:RISbeampattern}.
 
\section{Active RIS Beam Pattern Evaluation}
\label{sec:RISbeampattern}
\begin{figure}
	\centering
	\includegraphics[width=.9\columnwidth]{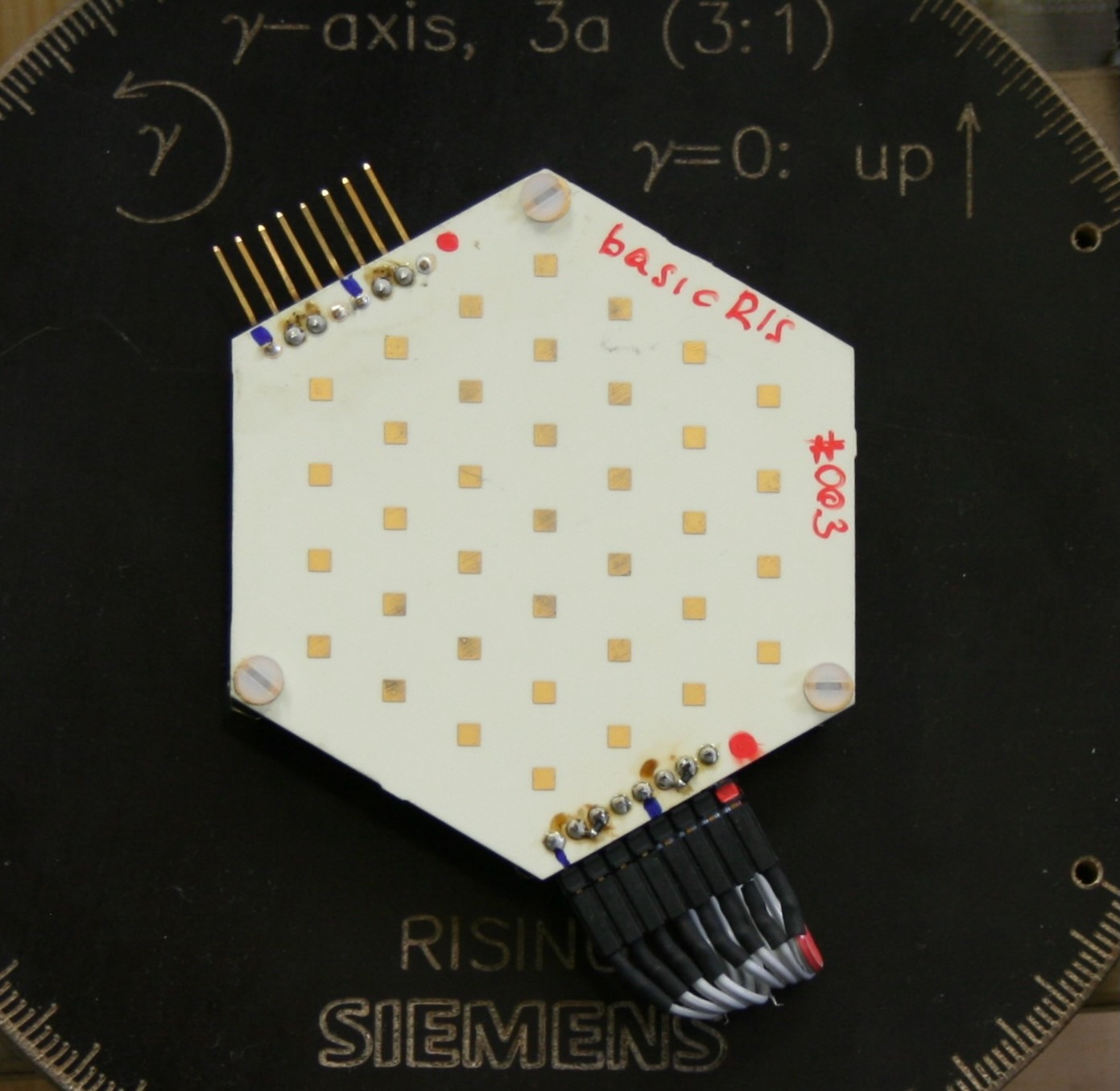}
	\caption{RIS PCB installed in the measurement setup.} 
	\label{fig:RIScloseup}
\vspace{-4mm}
\end{figure}

The manufactured active RIS PCB is shown in Fig. \ref{fig:RIScloseup}. The $M=37$ RIS elements are arranged in three hexagonal rings and an additional element is at the center of the RIS.
We summarize the parameters of the active RIS and the measurement setup in Table \ref{tab:ActiveRIS}.
\begin{table}
\begin{center}
\caption{Active RIS Parameters and Measurement Setup.}
\label{tab:ActiveRIS}
\begin{tabular}{ll} 
\toprule
Parameter	      	&  Definition\\
\midrule
$f = 25.8$\,GHz		& 	 center frequency \\
$M = 37$    & number of RIS elements\\
$d_z$, $d_y = 6.6$\,mm	& 	 effective RIS element size \\
$d = 8.7$\,mm & smallest RIS element distance\\
\midrule
$P_\text{BS}=10$\,dBm		& 	 BS transmit power\\
$G_\text{BS},G_\text{UE} = 19$\,dB		&  BS und UE horn antenna gain\\
$|\Vec{a}|,|\Vec{b}|= 1.7$\,m & distance RIS-BS and RIS-UE\\
$\tilde{\Vec{a}}= ( 1.7\,\text{m}, -25^\circ, 0^\circ)$ & BS location\\
$\tilde{\Vec{b}}=( 1.7\,\text{m}, 15^\circ, 30^\circ)$ & UE location\\
\bottomrule
\end{tabular}
\end{center}
\end{table}

\subsection{Numerical Beam Pattern Simulation}
\begin{figure}
	\centering
	\includegraphics[width=\columnwidth]{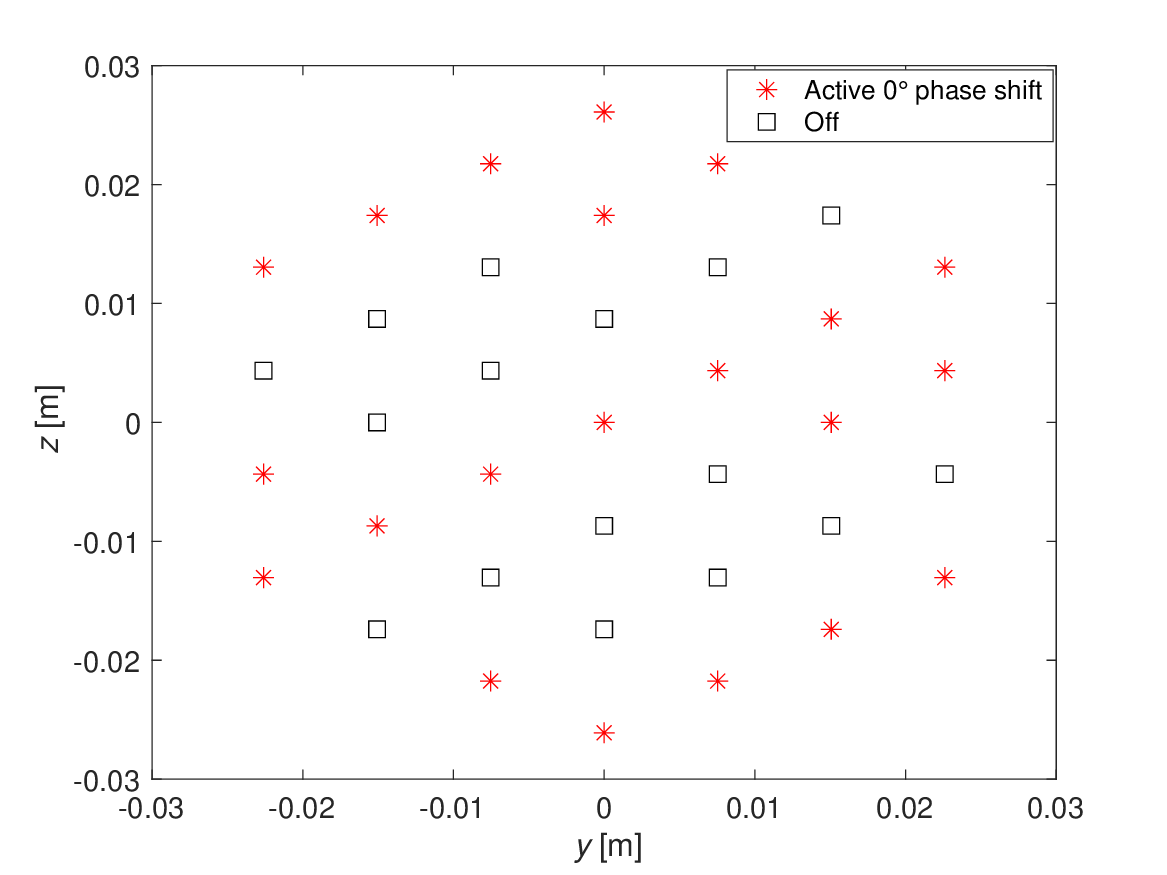}
	\caption{Active RIS element configuration using $\mathcal{A}_\text{A}$ for a desired main lobe focusing at the UE position $\tilde{\Vec{b}}= ( 1.7\,\text{m}, 15^\circ, 30^\circ)$ with the base station at $\tilde{\Vec{a}}= ( 1.7\,\text{m}, -25^\circ, 0^\circ)$. The red stars indicate active RIS elements (switched on), the rectangles indicate switched off RIS elements.} 
	\label{fig:RISconfig}
		\vspace{-4mm}
\end{figure}

We approximate \eqref{eq:MaximumReceivedPower} using Algorithm \ref{alg:MCsearch} for a specified position of the BS and the UE to obtain the RIS beam pattern numerically. For the reflective mode we use the reflection coefficient alphabet $\mathcal{A}_\text{R}$ and for the active mode we use $\mathcal{A}_\text{A}$. The resulting RIS element configuration for the active mode is depicted in Fig. \ref{fig:RISconfig}.

The beam pattern is obtained by evaluating (1) for the obtained RIS element configuration and a range of azimuth $-45 \leq b_\varphi \leq 45$ and elevation $-45 \leq b_\theta \leq 45$ values in $1^\circ$ steps. The numerical simulation results are shown for the reflective mode in Fig. \ref{fig:PassiveResult}(a) and for the active mode in Fig. \ref{fig:ActiveResult}(a).

\subsection{Empirical Beam Pattern Measurement}
\begin{figure}
	\centering
	\includegraphics[width=\columnwidth]{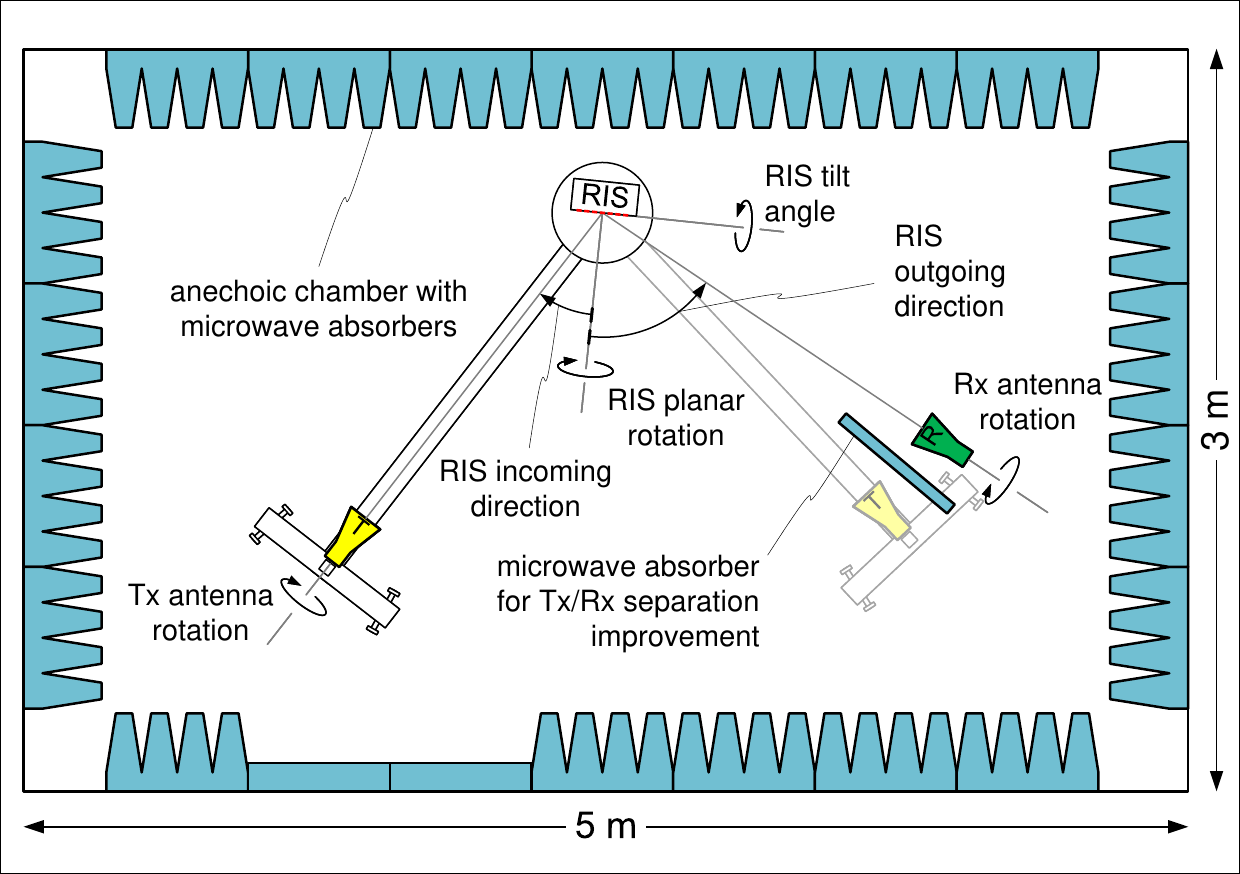}
	\caption{Measurement setup in anechoic chamber. The transmitter (TX) can be moved in azimuth and the receiver (RX) is fixed. Both can be rotated by $90^{\circ}$ to analyze vertical and horizontal polarization. The active RIS is mounted on a fixed position and can be rotated in all three axis.} 
	\label{fig:RISmeasurementschematic}
	\vspace{-4mm}
 \end{figure}

The beam pattern is measured in an anechoic chamber, using the setup shown in Fig. \ref{fig:RISmeasurementschematic}.
To minimize undesired reflected signals, microwave absorbers have been placed on all reflective parts of the measurement setup.

The motors of the measurement setup, the RIS element configuration, and the AIT mmWave channel sounder \cite{Hofer21} are controlled and synchronized by a computer system.  The channel sounder is used to obtain the channel impulse response for the BS-RIS-UE link. The sounding bandwidth is $B=155.5\,$MHz, resulting in a thermal noise power level of $N=-87\,$dBm taking the noise figure of the receiver of $5\,$dB into account. A calibration measurement is performed initially by connecting the cables from transmitter (TX) and receiver (RX) to the antenna foot points directly via an attenuator. With this calibration all effects of cables and amplifiers can be removed.

We perform also an initial measurement of the impulse response $h'_{\varphi,\theta}[n]$ vs. azimuth $\varphi$ and elevation $\theta$ with all elements of the RIS in off state. This impulse response represents all residual deterministic reflections of the anechoic chamber that we can subtract from later measurements of the RIS. Hence, we compute the received power
\begin{equation}
  P_\text{UE} (\varphi,\theta)=P_\text{BS}\sum_{n=n_1}^{n_2}\left \vert h_{\varphi,\theta}[n] - h'_{\varphi, \theta}[n]\right\vert^2.
\end{equation}
to obtain the beam pattern of the RIS. Here, $n_1$ and $n_2$ define the support of the impulse response above the noise floor to suppress measurement noise as much as possible. The impulse response measurements in this paper have a support in the interval $n_1\leq n \leq n_2$ with $n_1= 6$ and $n_2 = 10$. 

\begin{figure*}
	\centering
	\subfloat[Numerical simulation.]{\includegraphics[width=0.4\paperwidth]{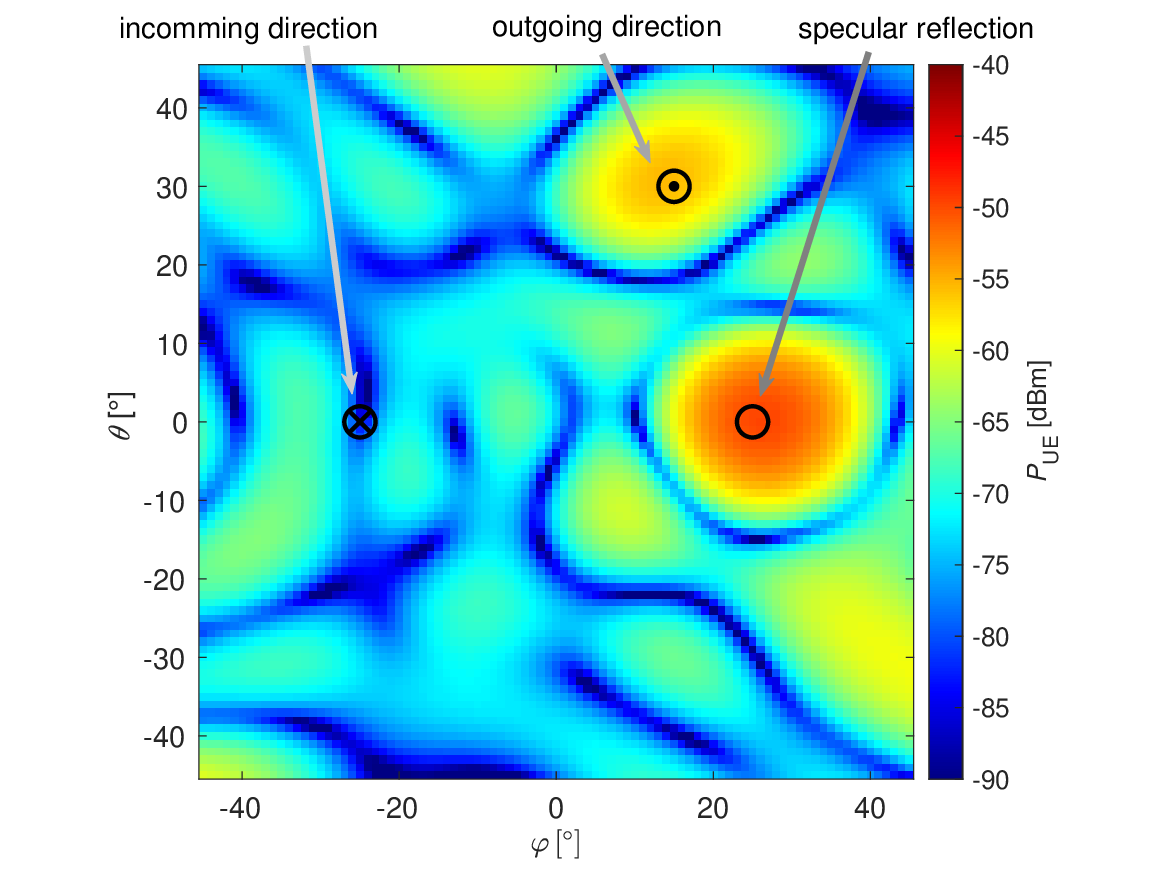}}
	\subfloat[Empirical measurements.]{\includegraphics[width=0.4\paperwidth]{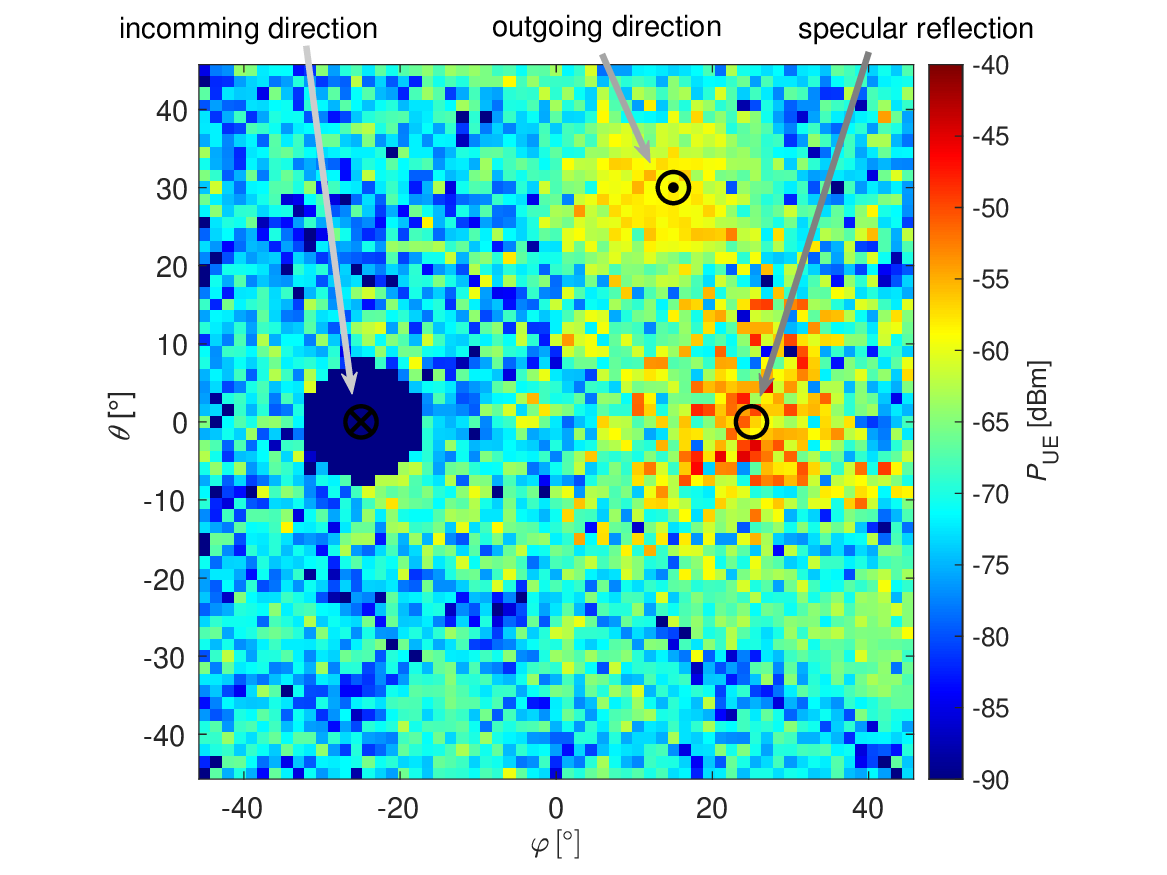}}
    \caption{Received power $P_\text{UE}(\varphi, \theta)$ vs. azimuth $\varphi$ and elevation $\theta$. We compare (a) numerical simulation results and (b) empirical measurement data for the RIS in reflective mode. The BS is located at $\tilde{\Vec{a}}= ( -25^\circ, 0^\circ, 1.7\,\text{m})$ and the signal shall be focused at the UE position $\tilde{\Vec{b}}=(15^\circ, 30^\circ, 1.7\,\text{m})$.}
	\label{fig:PassiveResult}
		\vspace{-4mm}
\end{figure*}

\begin{figure*}
	\centering
	\subfloat[Numerical simulation.]{\includegraphics[width=0.4\paperwidth]{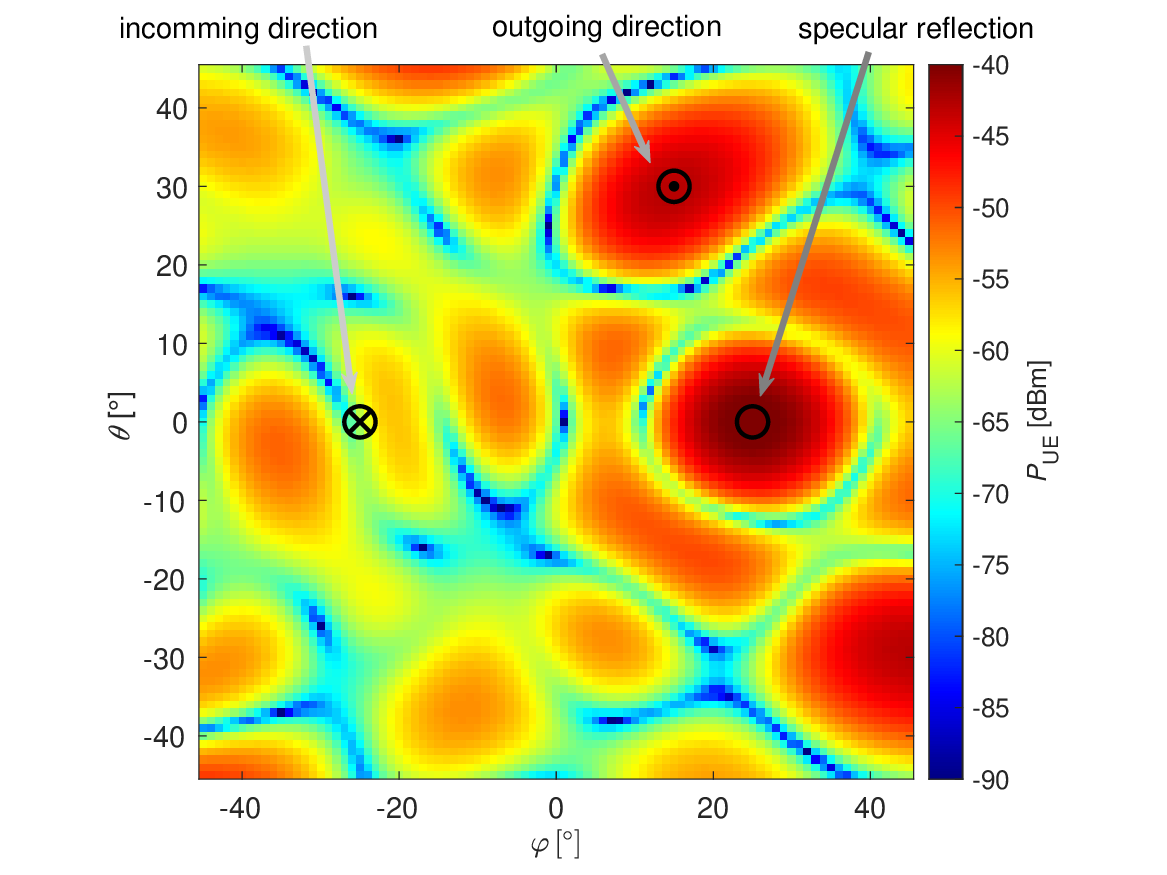}}
	\subfloat[Empirical measurements.]{\includegraphics[width=0.4\paperwidth]{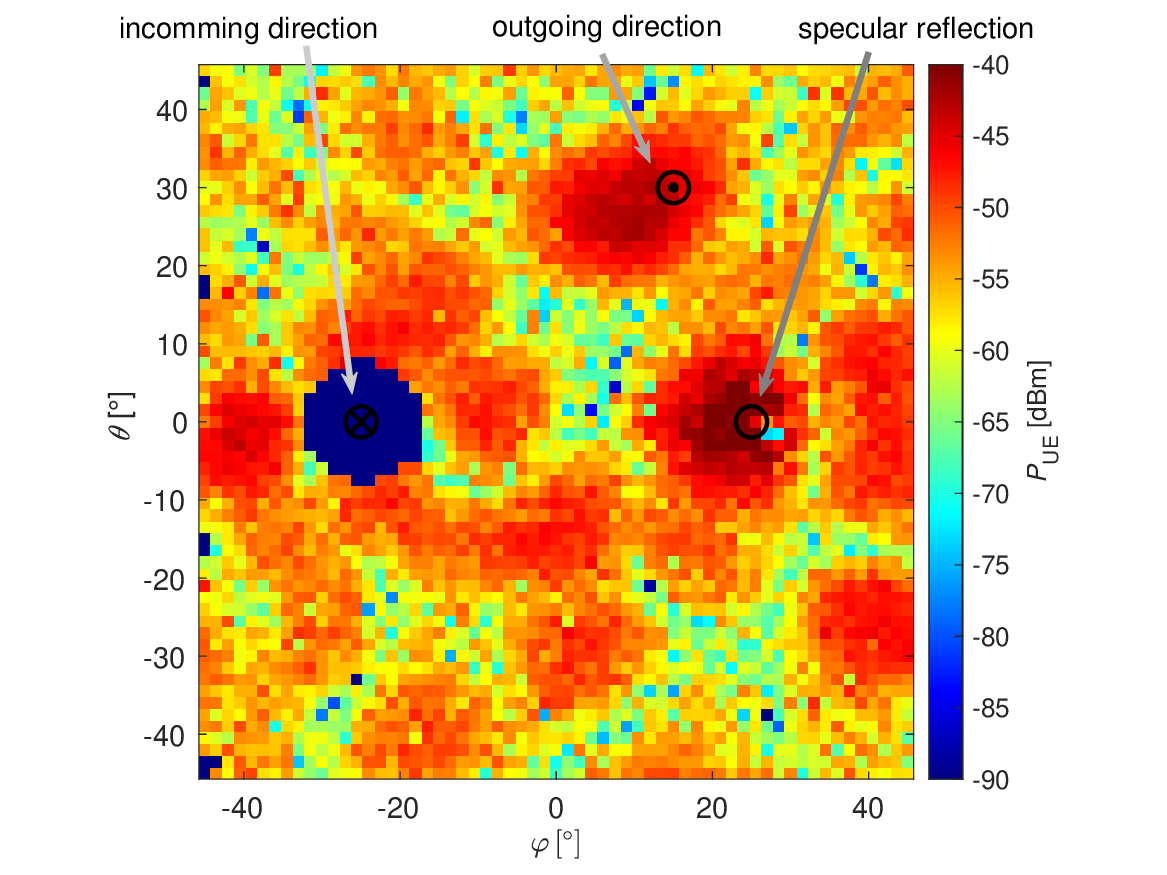}}
	\caption{Received power $P_\text{UE}(\varphi, \theta)$ vs. azimuth $\varphi$ and elevation $\theta$. We compare (a) numerical simulation results and (b) empirical measurement data for the RIS in active mode. The BS is located at $\tilde{\Vec{a}}= ( -25^\circ, 0^\circ, 1.7\,\text{m})$ and the signal shall be focused at the UE position $\tilde{\Vec{b}}=(15^\circ, 30^\circ, 1.7\,\text{m})$.}
	\label{fig:ActiveResult}
\vspace{-4mm}
\end{figure*}

Empirical measurement results are shown for the reflective mode in Fig. \ref{fig:PassiveResult}(b) and for the active mode in Fig. \ref{fig:ActiveResult}(b). Please note that the points close to the direction of the incoming signal cannot be measured, because the physical size of the transmit and receive antenna result in a minimum angle between transmit and receive antenna of $8^\circ$.

\section{Measurement Result Analysis} 
\label{sec:Comparison}
\subsection{Reflective Mode}
In reflective mode the RIS is able to steer the outgoing signal towards the desired direction, see Fig. \ref{fig:PassiveResult}. In both, simulation and measurement we obtained $P_\text{UE}(15^\circ, 30^\circ)= -55\, \text{dBm}$. The limited phase shift of $67^\circ$ has two effects: (i) The received power at the UE position is reduced by about $4\,$dB when compared to an ideal RIS with $180^\circ$ phase shift difference between both RIS element states (see Fig. \ref{fig:ReflectiveIdeal}) and (ii) the specular reflection of the BS signal at $(25^\circ, 0^\circ)$ is clearly visible in the beam pattern. For an ideal RIS the specular reflection would not be present as depicted in Fig. \ref{fig:ReflectiveIdeal}.

\subsection{Active Mode}
In active mode we can observe a stronger signal at the UE location, of about $P_\text{UE}(15^\circ, 30^\circ)=-43\, \text{dBm}$ for simulation and measurement results, see Fig. \ref{fig:ActiveResult}. In active mode the specular reflection is also clearly present. Note that for the active mode, elements can either be switched ON or OFF, which means that the group characteristic of the outgoing signal is not defined by a phase-shift at the RIS elements but rather by selecting and activating those RIS elements which contribute constructively at the UE position. According to the desired target angle, some elements (maximum half of the elements) must be switched OFF and do not contribute. To avoid the specular component an active RIS with a phase difference of $180^\circ$ between both RIS element states would be required, similar as in the reflective case above.

\section{Conclusion}
We presented an active RIS for the mmWave frequency band that can either operate in reflective mode or in active mode. In the active state the RIS elements use a polarization transform to increase the isolation between impinging and reflected wave. Up to the best of our knowledge, this is the first active RIS working at mmWave frequencies. We compare the reflective and active mode by means of numerical simulation and empirical measurements in an anechoic chamber showing a good match. Both, simulations and measurements indicate, that the active RIS is able to provide about $12\,$dB more received power at the UE position when compared with the reflective operation mode. 

\section*{Acknowledgment}
This work is funded through the Vienna Business Agency in the project RISING and by the Principal Scientist grant at the AIT Austrian Institute of Technology within project DEDICATE.


\end{document}

%% file: abbrev.tex
\begin{acronym} 
\setlength{\itemsep}{-0.63\parsep}
\acro{3GPP}{3rd Generation Partnership Project}
\acro{ACF}{auto correlation function}
\acro{ADC}{analog-to-digital converter}
\acro{BEM}{basis-expansion model}
\acro{CDF}{cumulative distribution function}
\acro{CE}{complex exponential}
\acro{CIR}{channel impulse response}
\acro{CP}[CP]{cyclic prefix}
\acro{CTF}{channel transfer function}
\acro{D2D}{device-to-device}
\acro{DAC}{digital-to-analog converter}
\acro{DC}{direct current}
\acro{DOD}{direction of departure}
\acro{DOA}{direction of arrival}
\acro{DPS}{discrete prolate spheroidal}
\acro{DPSWF}{discrete prolate spheroidal wave function}
\acro{DSD}{Doppler spectral density}
\acro{DSP}{digital signal processor}
\acro{DR}{dynamic range}
\acro{ETSI}{European Telecommunications Standards Institute}
\acro{FIFO}{first-input first-output}
\acro{FFT}{fast Fourier transform}
\acro{FP}{fixed-point}
\acro{FPGA}{field programmable gate array}
\acro{GSCM}{geometry-based stochastic channel model}
\acro{GSM}{global system for mobile communications}
\acro{GPS}{global positioning system}
\acro{HPBW}{half power beam width}
\acro{ICI}[ICI]{inter-carrier interference}
\acro{IDFT}{inverse discrete Fourier transform}
\acro{IF}{intermediate frequency}
\acro{IFFT}{inverse fast Fourier transform}
\acro{ISI}{inter-symbol interference}
\acro{ITS}{intelligent transportation system}
\acro{MEC}{mobile edge computing}
\acro{MSE}{mean square error}
\acro{LLR}{log-likelihood ratio}
\acro{LO}{local oscillator}
\acro{LOS}{line-of-sight}
\acro{LMMSE}{linear minimum mean squared error}
\acro{LNA}{low noise amplifier}
\acro{LSF}{local scattering function}
\acro{LTE}{long term evolution}
\acro{LUT}{look-up table}
\acro{LTV}{linear time-variant }
\acro{MIMO}{multiple-input multiple-output}
\acro{MPC}{multi-path component}
\acro{MC}{Monte Carlo}
\acro{NI}{National Instruments}
\acro{LoS}{line-of-sight}
\acro{NLOS}{non-line of sight}
\acro{OFDM}{orthogonal frequency division multiplexing}
\acro{OTA}{over-the-air}
\acro{PA}{power amplifier}
\acro{PC}{personal computer}
\acro{PDP}{power delay profile}
\acro{PER}{packet error rate}
\acro{PPS}{pulse per second}
\acro{QAM}{quadrature ampltiude modulation}
\acro{QPSK}{quadrature phase shift keying}

\acro{RB}{resource block}
\acro{RBP}{resource block pair}
\acro{RIS}{reflective intelligent surface}
\acro{RF}{radio frequency}
\acro{RMS}{root mean square}
\acro{RSSI}{receive signal strength indicator}
\acro{RT}{ray tracing}
\acro{RX}{receiver}
\acro{SCME}{spatial channel model extended}
\acro{SDR}{software defined radio}
\acro{SISO}{single-input single-output}
\acro{SoCE}{sum of complex exponentials}
\acro{SNR}{signal-to-noise ratio}
\acro{SUT}{system-under test}
\acro{SSD}{soft sphere decoder}
\acro{TBWP}{time-bandwidth product}
\acro{TDL}{tap delay line}
\acro{TX}{transmitter}
\acro{UMTS}{universal mobile telecommunications systems}
\acro{UDP}{user datagram protocol}
\acro{URLLC}{ultra-reliable and low latency communication} 
\acro{US}{uncorrelated-scattering}
\acro{USRP}{universal software radio peripheral}
\acro{VNA}{vector network analyzer}
\acro{ViL}{vehicle-in-the-loop}
\acro{V2I}{vehicle-to-infrastructure}
\acro{V2V}{vehicle-to-vehicle}
\acro{V2X}{vehicle-to-everything}
\acro{VST}{vector signal transceiver}
\acro{VTD}{Virtual Test Drive}
\acro{WF}{Wiener filter}
\acro{WSS}{wide-sense-stationary}
\acro{WSSUS}{wide-sense-stationary uncorrelated-scattering}
\end{acronym}

%% file: Radpour_WCNC2024.bbl
\begin{thebibliography}{10}
\providecommand{\url}[1]{#1}
\csname url@samestyle\endcsname
\providecommand{\newblock}{\relax}
\providecommand{\bibinfo}[2]{#2}
\providecommand{\BIBentrySTDinterwordspacing}{\spaceskip=0pt\relax}
\providecommand{\BIBentryALTinterwordstretchfactor}{4}
\providecommand{\BIBentryALTinterwordspacing}{\spaceskip=\fontdimen2\font plus
\BIBentryALTinterwordstretchfactor\fontdimen3\font minus
  \fontdimen4\font\relax}
\providecommand{\BIBforeignlanguage}[2]{{%
\expandafter\ifx\csname l@#1\endcsname\relax
\typeout{** WARNING: IEEEtran.bst: No hyphenation pattern has been}%
\typeout{** loaded for the language `#1'. Using the pattern for}%
\typeout{** the default language instead.}%
\else
\language=\csname l@#1\endcsname
\fi
#2}}
\providecommand{\BIBdecl}{\relax}
\BIBdecl

\bibitem{Basar19}
E.~Basar, M.~Di~Renzo, J.~De~Rosny, M.~Debbah, M.-S. Alouini, and R.~Zhang,
  ``Wireless communications through reconfigurable intelligent surfaces,''
  \emph{IEEE Access}, vol.~7, pp. 116\,753--116\,773, 2019.

\bibitem{khawaja2020coverage}
W.~Khawaja, O.~Ozdemir, Y.~Yapici, F.~Erden, and I.~Guvenc, ``Coverage
  enhancement for {NLOS} {mmWave} links using passive reflectors,'' \emph{IEEE
  Open Journal of the Communications Society}, vol.~1, pp. 263--281, 2020.

\bibitem{Rains23}
J.~Rains, J.~ur~Rehman~Kazim, A.~Tukmanov, T.~J. Cui, L.~Zhang, Q.~H. Abbasi,
  and M.~A. Imran, ``High-resolution programmable scattering for wireless
  coverage enhancement: An indoor field trial campaign,'' \emph{{IEEE} Trans.
  Antennas Propag.}, vol.~71, no.~1, pp. 518--530, 2023.

\bibitem{Dai20}
L.~Dai, B.~Wang, M.~Wang, X.~Yang, J.~Tan, S.~Bi, S.~Xu, F.~Yang, Z.~Chen,
  M.~D. Renzo, C.-B. Chae, and L.~Hanzo, ``Reconfigurable intelligent
  surface-based wireless communications: Antenna design, prototyping, and
  experimental results,'' \emph{IEEE Access}, vol.~8, pp. 45\,913--45\,923,
  2020.

\bibitem{Bjornson20a}
E.~Bj{\"o}rnson, {\"O}.~{\"O}zdogan, and E.~G. Larsson, ``Reconfigurable
  intelligent surfaces: Three myths and two critical questions,'' \emph{IEEE
  Communications Magazine}, vol.~58, no.~12, pp. 90--96, 2020.

\bibitem{najafi2020physics}
M.~Najafi, V.~Jamali, R.~Schober, and H.~V. Poor, ``Physics-based modeling and
  scalable optimization of large intelligent reflecting surfaces,''
  \emph{{IEEE} Trans. Commun.}, vol.~69, no.~4, pp. 2673--2691, 2020.

\bibitem{tang21}
W.~Tang, M.~Z. Chen, X.~Chen, J.~Y. Dai, Y.~Han, M.~Di~Renzo, Y.~Zeng, S.~Jin,
  Q.~Cheng, and T.~J. Cui, ``Wireless communications with reconfigurable
  intelligent surface: Path loss modeling and experimental measurement,''
  \emph{{IEEE} Trans. Wireless Commun.}, vol.~20, no.~1, pp. 421--439, 2021.

\bibitem{zhi2022active}
K.~Zhi, C.~Pan, H.~Ren, K.~K. Chai, and M.~Elkashlan, ``Active {RIS} versus
  passive {RIS}: Which is superior with the same power budget?'' \emph{IEEE
  Communications Letters}, vol.~26, no.~5, pp. 1150--1154, 2022.

\bibitem{Zhang23}
Z.~Zhang, L.~Dai, X.~Chen, C.~Liu, F.~Yang, R.~Schober, and H.~V. Poor,
  ``Active {RIS} vs. passive {RIS}: Which will prevail in {6G}?'' \emph{IEEE
  Transactions on Communications}, vol.~71, no.~3, pp. 1707--1725, 2023.

\bibitem{rao23}
J.~Rao, Y.~Zhang, S.~Tang, Z.~Li, C.-Y. Chiu, and R.~Murch, ``An active
  reconfigurable intelligent surface utilizing phase-reconfigurable reflection
  amplifiers,'' \emph{IEEE Transactions on Microwave Theory and Techniques},
  vol.~71, no.~7, pp. 3189--3202, 2023.

\bibitem{wu2022wideband}
L.~Wu, K.~Lou, J.~Ke, J.~Liang, Z.~Luo, J.~Y. Dai, Q.~Cheng, and T.~J. Cui, ``A
  wideband amplifying reconfigurable intelligent surface,'' \emph{{IEEE} Trans.
  Antennas Propag.}, vol.~70, no.~11, pp. 10\,623--10\,631, 2022.

\bibitem{tang22}
W.~Tang, X.~Chen, M.~Z. Chen, J.~Y. Dai, Y.~Han, M.~Di~Renzo, S.~Jin, Q.~Cheng,
  and T.~J. Cui, ``Path loss modeling and measurements for reconfigurable
  intelligent surfaces in the millimeter-wave frequency band,'' \emph{{IEEE}
  Trans. Commun.}, vol.~70, no.~9, pp. 6259--6276, 2022.

\bibitem{Hofer21}
M.~Hofer, D.~L{\"o}schenbrand, J.~Blumenstein, H.~Groll, S.~Zelenbaba,
  B.~Rainer, L.~Bernad{\'o}, J.~Vychodil, T.~Mikulasek, E.~Z{\"o}chmann,
  S.~Sangodoyin, H.~Hammoud, B.~Schrenk, R.~Langwieser, S.~Pratschner,
  A.~Prokes, A.~F. Molisch, C.~F. Mecklenbr{\"a}uker, and T.~Zemen, ``Wireless
  vehicular multiband measurements in centimeterwave and millimeterwave
  bands,'' in \emph{IEEE International Symposium on Personal, Indoor and Mobile
  Radio Communications (PIMRC)}, virtual conference, 2021, pp. 836--841.

\end{thebibliography}
